\documentclass[sigconf,screen]{acmart}

\AtBeginDocument{%
  \providecommand\BibTeX{{%
    Bib\TeX}}}

\def\BibTeX{{\rm B\kern-.05em{\sc i\kern-.025em b}\kern-.08em
    T\kern-.1667em\lower.7ex\hbox{E}\kern-.125emX}}
  
\usepackage{xcolor}
\usepackage{graphicx}
\usepackage{subcaption}
\usepackage{amsmath}
\usepackage{array}

\usepackage{tcolorbox}
\usepackage{xspace}
\usepackage{enumitem}
\usepackage{multirow}
\newcolumntype{C}[1]{>{\centering\arraybackslash}m{#1}}
\usepackage{tabularx}
\usepackage[T1]{fontenc}

\newcommand{\cmmnt}[1]{}

\usepackage{microtype}
\newcommand{\todo}[1]{\textcolor{black}{#1}}
\newcommand{\todohs}[1]{\textcolor{black}{#1}}
\newcommand{\todob}[1]{\textcolor{black}{#1}}
\newcommand{\todop}[1]{\textcolor{black}{#1}}
\newcommand{\todoq}[1]{\textcolor{black}{#1}}
\newcommand{\todocr}[1]{\textcolor{black}{#1}}
\newcommand{\tododel}[1]{\textcolor{pink}{}}

\copyrightyear{2023} 
\acmYear{2023} 
\setcopyright{acmlicensed}
\acmConference[ESEC/FSE '23]{Proceedings of the 31st ACM Joint European Software Engineering Conference and Symposium on the Foundations of Software Engineering}{December 3--9, 2023}{San Francisco, CA, USA}
\acmBooktitle{Proceedings of the 31st ACM Joint European Software Engineering Conference and Symposium on the Foundations of Software Engineering (ESEC/FSE '23), December 3--9, 2023, San Francisco, CA, USA}
\acmPrice{15.00}
\acmDOI{10.1145/3611643.3616325}
\acmISBN{979-8-4007-0327-0/23/12}

\begin{document}

\title{Demystifying Dependency Bugs in Deep Learning Stack}

\author{Kaifeng Huang}
\authornote{K. Huang, B. Chen, S. Wu, J. Cao, and X. Peng  are with the School of Computer Science and Shanghai Key Laboratory of Data Science, Fudan University, China.}
\affiliation{
\institution{Fudan University}
\country{China}
}

\author{Bihuan Chen}
\authornotemark[1]
\authornote{B. Chen is the corresponding author.}
\affiliation{
\institution{Fudan University}
\country{China}
}

\author{Susheng Wu}
\authornotemark[1]
\affiliation{
\institution{Fudan University}
\country{China}
}

\author{Junming Cao}
\authornotemark[1]
\affiliation{
\institution{Fudan University}
\country{China}
}

\author{Lei Ma}
\authornote{L. Ma is also with University of Alberta, Canada.}
\affiliation{
\institution{The University of Tokyo}
\country{Japan}
}

\author{Xin Peng}
\authornotemark[1]
\affiliation{
\institution{Fudan University}
\country{China}
}

\renewcommand{\shortauthors}{K. Huang et al.}

\begin{abstract}
Deep learning (DL) applications, built upon~a~heterogeneous and complex DL stack (e.g., Nvidia GPU, Linux,~CUDA driver,~Python runtime, and TensorFlow), are subject to software and hardware dependencies~across~the DL stack. One~challenge~in~dependency management~across the entire engineering lifecycle~is posed by~the~asynchronous~and~radical evolution and  the complex version~constraints among~dependencies.~Developers may introduce dependency bugs (DBs) in selecting, using and~maintaining dependencies. However,~the characteristics of DBs in DL stack is still under-investigated, hindering practical solutions to dependency management in DL stack.


To bridge this gap, this paper presents the first comprehensive study to characterize symptoms, root causes and fix~patterns~of DBs~across the whole DL stack with \cmmnt{\todo{326}\todob{+120}~}\todob{446} DBs collected from StackOverflow~posts \todob{and GitHub issues}. For each DB, we~first investigate the symptom as well as the lifecycle stage and~dependency where the symptom is exposed.~Then,~we~analyze the root cause as well as the lifecycle stage and~dependency~where the root cause is introduced. Finally, we explore the fix pattern~and~the~knowledge~sources that are used to fix it. Our findings~from~this study shed light on practical implications~on~dependency management.


\end{abstract}



\maketitle


\section{Introduction}\label{sec:intro}

The significant breakthroughs in deep learning~(DL) have brought great success to many DL-enabled applications,~e.g., machine translation~\cite{JeanCMB15}, medical diagnosis~\cite{obermeyer2016predicting}, voice assistants~\cite{hauswald2015sirius}~and autonomous vehicles~\cite{chen2015deepdriving}. Such~DL~applications~are~built~upon~a heterogeneous and complex DL~stack, including hardware~(e.g.,~Nvidia GPU), OS (e.g., Linux),~drivers (e.g., CUDA and cuDNN),~runtime~(e.g., Python) and libraries (e.g., TensorFlow). In other words, engineering DL applications requires software and hardware~in~the DL~stack~as prerequisite dependencies. One common challenge in engineering DL applications~is~dependency management~across~the DL~stack~\cite{chen2020comprehensive, zhang2019empirical, alshangiti2019developing}, i.e.,~to properly manage versions and configurations of~the software and hardware dependencies in the entire~DL~stack. 

\textbf{Motivation.} Dependency management is challenging for~three main reasons. First, \textit{software and hardware dependencies are~complex, and evolve quickly in an asynchronous~and~radical way.} Dependency complexity originates from two sources,~i.e., deep~stack~and~rich vendors. For example, many vendors~provide DL~libraries,~e.g.,~Google's TensorFlow, Facebook's~PyTorch and Microsoft's CNTK.~Besides, dependency evolution~is performed at the vendor's own pace~and~may introduce~incompatible changes. For example, the micro-architecture of Nvidia GPU~has~evolved several generations over the years,~from old versions such~as~Tesla to new versions such as Ampere.~In the meantime, CUDA has evolved from version~1.0~to~11.6~to support different GPUs distinguished by compute capability, which ranges from 1.0 to 9.0. Therefore, developers~may~miss some dependencies and build an incomplete stack, or have~troubles in~selecting, updating and migrating dependency versions.

Second, \textit{software and hardware dependencies need to satisfy complex version constraints to work together properly.} For example,~each TensorFlow version only works~compatibly with certain~cuDNN~versions, CUDA versions~and~Nvidia~GPU~versions. A~developer set up an environment with~TensorFlow gpu version 1.2.0rc0,~Python 3.5.2, CUDA 8.0.61, cuDNN 8.0 and a GPU card with compute capability~2.1~on~Windows 7 \cite{gpu-capability}. The setup failed to recognize~a~valid~GPU card~as~this TensorFlow version required a GPU card with~compute~capability 3.0 or higher. 
These version constraints are scattered~across documentations of software and hardware. Therefore, developers may build an incompatible stack, or introduce~incompatibilities when updating versions or deploying to a new environment.


Third, \textit{each dependency version may contain bugs or need proper~configuration.} While dependency version constraints are satisfied,~there might be bugs in specific versions under~certain circumstances.~For example, a developer created~a~Seq2Seq model using TensorFlow~1.5 but encountered an error \cite{tensorflow-bug}.~It was caused by a bug~only in TensorFlow 1.5, and could~be~alleviated by upgrading to~1.6~or~downgrading to 1.4. In addition,~there might be misconfigurations during the installation of dependencies. For example, some kernel modules are required~to~be signed on Secure Boot enabled systems when the Nvidia~driver is installed. However, this may cause unknown errors raised from CUDA \cite{disable-secure-boot}, which could be fixed by disabling Secure Boot. Therefore, developers might use a buggy dependency~version~or~misconfigure a dependency version.


In summary, developers may introduce various dependency management problems in selecting, using and maintaining~dependencies in the DL stack during~the~entire engineering~lifecycle (i.e.,~environment setup, development, deployment~and~maintenance). We refer to these~problems as dependency bugs (DBs).

\begin{figure*}[!t]
    \centering
    \includegraphics[width=0.88\textwidth]{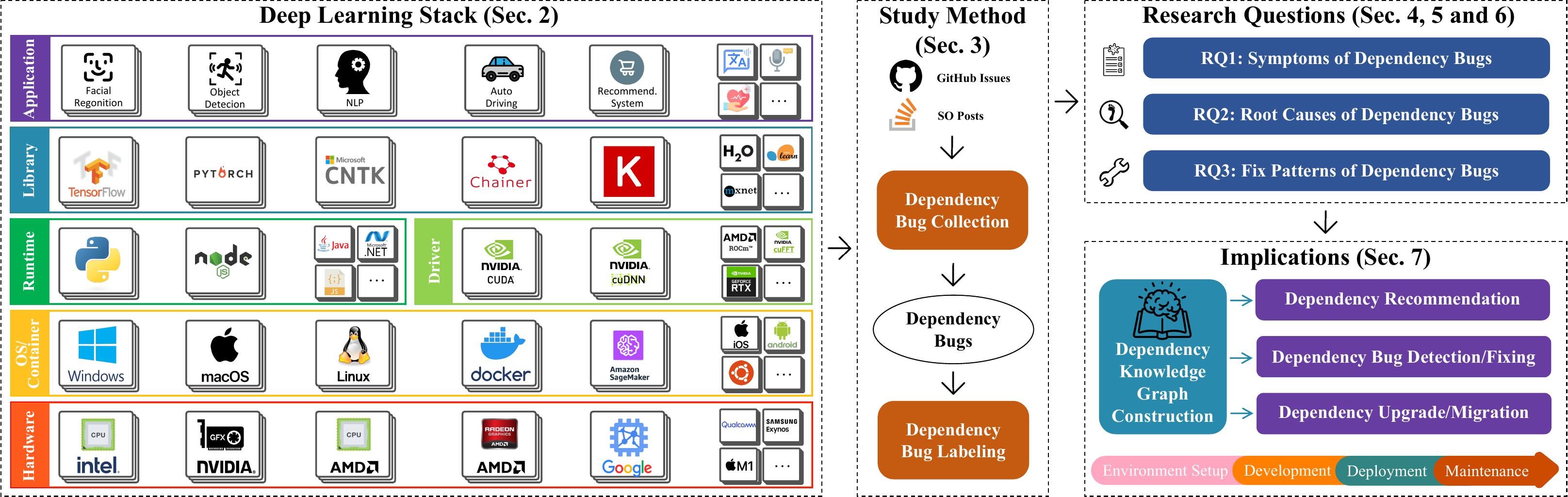}
    \vspace{-7pt}
    \caption{An Overview of Our Empirical Study on Dependency Bugs in DL Stack}
    \label{fig:overview}
\end{figure*}

\textbf{Literature.} On the one hand, a lot of advances have~been made to investigate~DBs in different ecosystems, e.g., Java~\cite{wang2018dependency, huang2020interactive}, C/C++~\cite{jia2021depowl}, JavaScript~\cite{patra2018conflictjs}, Python~\cite{mukherjee2021fixing, wang2020watchman}, Go~\cite{wang2021hero},~and Debian~and Red~Hat~\cite{artho2012software}. They only consider DBs~at~the~homogeneous library layer. However, DBs in the DL ecosystem are different because they can occur~across all the heterogeneous layers in the DL stack.~On~the~other~hand,~a~lot of efforts~have been made to investigate characteristics (e.g,~symptoms,~root causes and fix patterns) of general bugs~\cite{EmpiricalStudyTensorFlow2018, islam2019comprehensive, humbatova2019taxonomy, islam2020repairing, nikanjam2022faults} and specific bugs~\cite{zhang2020empirical, chen2021empirical, zhang2021autotrainer, MLAPI2021, cao2021characterizing} in DL applications.~However, these studies are not specifically designed for DBs, and thus only uncover partial characteristics of DBs in DL stack. 
Therefore, although it~is~necessary to understand the characteristics of DBs in DL stack, no systematic study exists yet. 





\textbf{Our Study.} To bridge this gap, we present the first~comprehensive study to characterize DBs in DL stack.~An~overview of~our study is presented in Fig.~\ref{fig:overview}. After introducing the~DL~stack (see Sec.~\ref{sec:stack}), we first collect \cmmnt{\todo{326}\todob{+120}~}\todob{446} DBs from~StackOverflow posts~\todob{and GitHub issues}, and~then analyze these DBs to answer three RQs (see Sec.~\ref{sec:study-design}).

\begin{itemize}[leftmargin=*]
    \item \textbf{RQ1 Symptom:} What are the symptoms of DBs?~At~which lifecycle stages and dependencies are they exposed?
    \item \textbf{RQ2 Root Cause:} What are the root causes of DBs?~At which lifecycle stages and dependencies are they introduced?
    \item \textbf{RQ3 Fix Pattern:} What are the fix patterns of DBs?~Which knowledge sources are used to fix DBs?
\end{itemize}

Through these research questions, we aim to provide~useful findings for developers and researchers (see Sec.~\ref{sec:symptom}, \ref{sec:root-cause}~and~\ref{sec:fix-pattern}). For example, \cmmnt{\todo{36.2\%},118/326, 118+55 / 446~}\todob{38.8\%} of the DBs manifest DL~specific errors or anomalies~in~software and hardware dependencies,~behavior, model and data, mostly~leading to crashes. Violation~of~constraints among software and hardware dependencies~causes \cmmnt{\todo{85.3\%}, 278/326, 356/446~}\todob{79.8\%} of the DBs. Development is the most bug-affecting~lifecycle stage, which exposes~\cmmnt{\todo{61.3\%}, 200/326, 231/446~}\todob{51.8\%} of the DBs, while environment setup is the most bug-prone lifecycle stage, which~introduces \cmmnt{\todo{87.4\%}, 285/326, 405/446~}\todob{90.8\%} of the DBs. \cmmnt{\todo{43.6\%}, 142/326 }\todob{227 (50.9\%)} of the DBs are~not~introduced and~exposed in the same dependency. Changing dependency version and adding dependency~are the most common fix~patterns, which are leveraged to fix \cmmnt{\todo{78.5\%}, 256/324, 312/446~}\todob{70.0\%} and \cmmnt{\todo{12.6\%},41/326, 43/446~}\todob{11.9\%} of the DBs. 
\todop{Source code, documentation, issue~tracker~and other online resource are important knowledge sources~of~fixing DBs.}

Our findings provide practical implications for developers and researchers~on dependency management across the entire engineering lifecycle~(see Sec.~\ref{sec:implication}), e.g., construct dependency knowledge graph for the entire DL stack, recommend dependencies~in~the entire DL stack, detect, localize and fix dependency bugs, and upgrade and migrate dependencies. \todo{To demonstrate the usefulness of our implications, we design~a~prototype~of~DB detection and fixing.}

In summary, our work makes the following contributions.

\begin{itemize}[leftmargin=*]
    \item We conduct the first comprehensive study to explore~symptoms, root causes and fix patterns of \cmmnt{\todo{326}\todob{+120}~}\todob{446} DBs~in~DL~stack.
    \item We provide implications for developers and researchers on dependency management in engineering~DL~applications.
\end{itemize}


\section{Deep Learning Stack}\label{sec:stack}

Developers need to set up a DL environment before developing~or~deploying DL applications. The setup process often involves the following steps. First, developers need to choose a physical~machine with GPUs and operating system installed. Besides, developers~can use a virtual machine on the physical machine, or choose a virtual machine on the cloud supported by cloud service providers~(e.g., Amazon SageMaker). Second, to fully empower upper libraries~and DL applications, developers need to install the corresponding GPU drivers and GPU-accelerated SDKs (e.g., CUDA and cuDNN). Third, developers need to select a runtime environment based~on~the~programming language that DL applications are developed with (e.g., Python and Java). Forth, a number of libraries should be leveraged to boost the development of DL applications from different perspectives. Finally, developers could develop and deploy DL applications on top of the software and hardware dependencies. 






This setup process is complicated by involving a wide~scope~of software and hardware dependencies. To reduce the complexity~and provide a complete solution, a DL stack is proposed~by organizing dependencies into layers.~For example, Patterson shows~a~generic program stack consisting of modeling code, framework, storage,~driver, operating system and hardware~\cite{josh-stack}. By following the setup~process and referencing~the~DL~stack~at Patterson Consulting~\cite{josh-stack},~Intel~\cite{intel}, Huawei~\cite{huawei} and Nvidia \cite{docker-gpu},  we summarize a DL stack~in Fig.~\ref{fig:overview}. It consists of five layers. From top to bottom, they are~\textit{Application}, \textit{Library}, \textit{Runtime} and \textit{Driver}, \textit{OS/Container}, and \textit{Hardware}. 

Specifically, the \textit{Application} layer contains DL applications from various domains, e.g., autonomous driving. The \textit{Library} layer contains the dependencies~the~upper-layer DL~applications directly or transitively depend on. It covers a wide range~of libraries,~including frameworks (e.g., TensorFlow, PyTorch~and~CNTK) which provide abstraction~and~generic functionality implementation for DL algorithms,~front-end~libraries providing high-level~abstraction or language bindings (e.g., Keras, ktrain and NeuPy),~and other libraries in the~ecosystem. The \textit{Runtime} layer includes interpreters for dynamically typed languages (e.g., Python~and JavaScript)~and~virtual~machines for statically typed languages (e.g., Java and .Net). The \textit{Driver} layer contains the dependencies for interacting with GPUs, including GPU drivers, computing platforms and GPU-accelerated SDKs (e.g., Nvidia GPU driver, CUDA and cuDNN).~The \textit{Library} layer can directly interact with the \textit{Runtime} and \textit{Driver}~layer, and thus they are put at the same layer. The \textit{OS/Container} layer~contains operating systems, containers and other virtual environments (e.g., Ubuntu, Windows, macOS, Docker, and Amazon SageMaker). The \textit{Hardware} layer contains fundamental hardware like CPU, GPU, mobile chips, and vendor-specific chips (e.g., Google's TPU). 



\section{Empirical Study Methodology}\label{sec:study-design}

We first introduce the design of our empirical study, and then~present our process of data collection and data labeling.

\subsection{Study Design}


Our \textit{symptom analysis} in \textbf{RQ1} aims to characterize~the~observable consequences of DBs, which is helpful~to~assess~impacts and~provide insights for DB diagnosis and detection. Moreover, it aims~to~identify the lifecycle stage and dependency where the symptom~is exposed, which is helpful to guide both developers and researchers~to~focus more effort on these bug-affecting stages and dependencies~so~as~to achieve the most benefit for DB diagnosis and detection. 

Our \textit{root cause analysis} in \textbf{RQ2} seeks to understand the fundamental nature of DBs, which is helpful to provide~insights~for~DB detection and localization. Further, it seeks~to~locate~the lifecycle stage and dependency where the root cause~is~introduced, which~is~helpful to guide both~developers~and~researchers to~spend more effort~on these bug-prone stages and dependencies in order to achieve the most benefit for DB avoidance, detection and localization. 

Our \textit{fix pattern analysis} in \textbf{RQ3} attempts to characterize~the fixes of DBs, which is helpful to provide insights for DB~fixing. Moreover, it explores the distribution of fix~patterns~for~root causes~as well as the knowledge sources that are used~to~fix DBs, which~is helpful for both developers and researchers to achieve DB fixing in a more automated and effective fashion.

\todoq{\textbf{Comparison to DBs in Other Domains.} 
Unlike general programming, DBs in deep learning exhibits a higher prevalence~of~low-level issues, e.g., driver configuration problems. To the best of our knowledge, there is no literature on DBs in high-performance computing or platform-specific binaries, which also encounter configuration problems that may be as prevalent as those found~in~deep learning. The existing literature covers a range of topics related to dependencies, including empirical studies on dependency smells~\cite{cao2022towards, jafari2021dependency}, dependency conflicts~\cite{artho2012software, patra2018conflictjs,wang2018dependency, huang2020interactive, wang2019could, wang2021will, wang2020watchman, wang2021hero} and dependency-related build failures~\cite{macho2018automatically, mukherjee2021fixing,bezemer2017empirical, wu2022accelerating, lou2020understanding}. During the analysis of \textbf{RQ1}, \textbf{RQ2} and \textbf{RQ3}, we compare the symptoms, root causes and fix patterns and discuss the differences from existing literature.}

\subsection{Data Collection}\label{empirical-setup}

\todob{To obtain a comprehensive understanding of DBs, we collect relevant posts on StackOverflow and relevant issues on GitHub. We selected StackOverflow and GitHub because i) they are popular sites containing a wide range~of~problems raised by world-wide developers in real-life development~activities; and ii) they have a high potential to contain problems about~the dependencies in the entire DL stack due to their diversity.}

\subsubsection{\todob{Collecting SO Posts}}

\todob{Our~collection of SO posts has two steps.}

\textbf{Step 1: Dependency Tag Selection.} Developers often~attach several tags to a post to indicate the topics or concepts~related~to~the question. Therefore, tags can be used to select the~posts that~are~relevant to dependency problems in DL stack, and we need to determine~a~set of tags that have a high coverage~of the dependencies in DL stack.~To this end, we first collected all the \todo{21,978,327}  posts from Stack~Exchange~Data Dump on December 20, 2021. Then,~for each post with an accepted answer, we iterated its tag~list,~and~searched for tags that co-occurred with the tag ``deep learning'' or ``neural network''. In this way, we obtained an initial set of \todo{1,576}~tags.~We~did~not directly use the tag ``deep learning'' or ``neural network''~to~select posts as it may miss posts that were not tagged~with~``deep learning'' and ``neural network'' but with other dependency related tags.



Next, two of the authors independently determined~whether each of the \todo{1,576} tags was related to the dependencies in DL stack~by~reading the excerpt provided by StackOverflow and online materials obtained by search engines. We used Cohen’s Kappa~coefficient~to~measure agreement, and it reached \todo{0.906}. A third~author~was~involved to~resolve disagreements. Finally, we obtained \todo{57} \textit{Library} tags,~\todo{3}~\textit{Driver} tags, \todo{59} \textit{Runtime} tags, \todo{23} \textit{OS/Container} tags and \todo{14} \textit{Hardware}~tags.

\todoq{Moreover, we conducted a comprehensive analysis of these 156 tags on significance and relevance scores, following previous~work \cite{bagherzadeh2019going, ahmed2018concurrency}. Out of these tags, 106 of them have non-zero scores in terms of both significance and relevance, while the remaining 50 tags have zero scores. These tags exhibit an average significance~score of 0.040 and an average relevance score of 0.018. Compared with the thresholds used in previous work \cite{bagherzadeh2019going, ahmed2018concurrency}, our results suggest that our set of DL stack tags is significant and relevant.}

\textbf{Step 2: Dependency Post Selection.} We picked dependency-related posts in two steps. First, we chose from~the~\todo{21,978,327} posts the ones whose tags contained one of the \todo{57} \textit{Library} tags and \todo{3}~\textit{Driver} tags, or contained the tag ``deep learning'' or ``neural network''~as well as one of the \todo{59} \textit{Runtime} tags, \todo{23} \textit{OS/Container} tags and \todo{14}~\textit{Hardware}~tags. As \textit{Runtime}, \textit{OS/Container} and \textit{Hardware} tags often have a weaker correlation with DL than \textit{Library} and \textit{Driver} tags,~here~we enforced their co-occurrence with either ``deep learning'' or ``neural network'' to reduce noisy posts. This led to \todo{66,422} posts.

Second, to focus on high-quality posts, \todoq{we removed \todocr{7,301} posts~that did not have an accepted answer and \todocr{35,327 posts} that did not contain dependency~version information}. The information of dependency versions~was~considered as important to determine root causes and fix patterns~of~DBs. We used regular expression matching to check the existence of version information. This restricted our selection to \todo{3,814} posts.

\subsubsection{\todob{Collecting GitHub Issues}}

\todob{Our collection of GitHub~issues consists of  two steps. }

\todob{\textbf{Step 1: GitHub Repsitory Selection.}~To obtain dependency-related issues, we need to select a set of repositories across~the DL stack. However, GitHub mainly hosts repositories~at~the~\textit{Application} and \textit{Library} layer. Therefore, we~first~searched~the~\todo{57} \textit{Library} tags in GitHub, which linked to 30 GitHub repositories. The repository size is smaller than the tag size~as~i)~some tags share the same~repository; ii) some~repositories~are archived; and iii) some libraries are not hosted on GitHub.~Then,~we~selected the top 10 repositories~in the \textit{Application} layer by querying GitHub using ``deep learning''.}

\todob{\textbf{Step 2: Dependency Issue Collection.}~We collected~closed issues in the 40 selected repositories using GitHub API, which led~to 154,299 issues. Similar to dependency post selection,~we used regular expression matching to check~the~existence~of~version information in issues, which resulted in 37,795 issues. As the issue size is still large, we randomly sampled 1,763 issues with a confidence level of 99\% and a margin of error~of~3\%.}

\subsubsection{\todob{DB Identification}}

\todob{We manually verified the~\todo{3,814}~posts and \todo{1,763 issues} to reduce noise that was not about~DBs~in~DL stack. 
In particular, two of the authors independently investigated each post and issue to identify DBs. The~Cohen's~Kappa coefficient was \cmmnt{\todo{0.899}~}\todob{0.0.909}. A third author was involved to resolve disagreements. Finally, we identified \todob{446} DBs. 326 are from posts, and 120 are from issues.}

\subsection{Data Labeling} 

To answer the three research questions, we manually labeled each~of the \cmmnt{\todo{326}~}\todob{446} DBs with respect to eight~aspects, i.e., symptom, exposing stage and dependency, root~cause,~introducing stage and dependency, fix pattern, and knowledge source for fixing. 

In particular, two of the authors first randomly sampled~\todo{100} DBs for a pilot labeling, following an open coding procedure \cite{seaman1999qualitative}. They separately read all contents~of~a~post~or~issue~(including title, question/issue~description, comments,~answers, commits and reference links mentioned during discussion) and relied~on search engines to carefully label~DBs. Basically,~the symptom of a DB was determined by analyzing~the~question/issue description. The root cause, fix pattern and knowledge source~for fixing of a DB were inferred from~the~question/issue description, the fixing commit or the accepted answer. The exposing stage and dependency~of~a~DB were determined by analyzing where its symptom was exhibited,~while the introducing stage and dependency of a DB were~determined~by~analyzing where its root cause was located. A group discussion~was~conducted to summarize the initial taxonomies.



Then, two of the authors independently labeled all the \cmmnt{\todo{326}~}\todob{446} posts based on the initial taxonomies, and finally reached~Cohen's~Kappa coefficients of \cmmnt{\todo{0.905, 0.822, 0.806, 0.767, 0.993, 0.798, 0.961 and 0.856~}}\todob{0.967, 0.938, 0.930, 0.840, 0.870, 0.887,~0.813 and~0.858~} for the eight~aspects.~A~third~author resolved disagreements in pilot and final labeling. The manual effort,~involved in our data collection and labeling, required \cmmnt{\todo{six}~}\todob{eight} person-months.


%
%
%
%
%
%
%


\section{RQ1: Symptom Analysis}\label{sec:symptom}

We present the taxonomy of DB symptoms, and explore the stages and dependencies where symptoms are exposed.

\subsection{Symptom Taxonomy}

\begin{figure}[!t]
    \centering
    \includegraphics[width=0.42\textwidth]{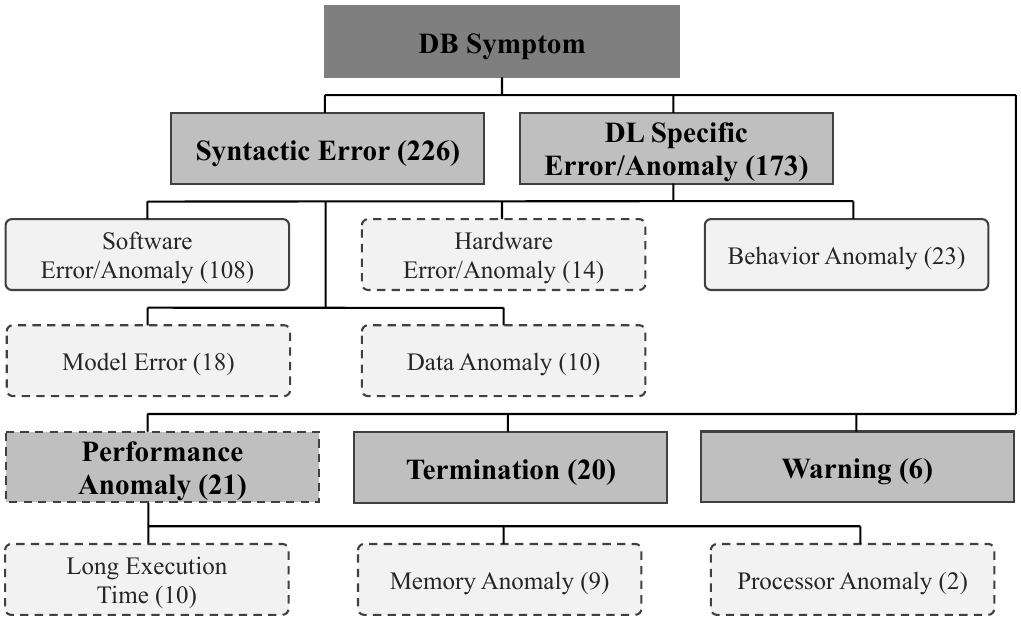}
    \vspace{-5pt}
    \caption{\todoq{Taxonomy of DB Symptoms}}
    \label{fig:error}
\end{figure}

The taxonomy of DB symptoms is reported in Fig.~\ref{fig:error}.~It~is organized into five inner categories (i.e. \textit{Syntactic Error},~\textit{DL~Specific Error/Anomaly}, \textit{Performance Anomaly}, \textit{Termination}~and \textit{Warning})~and \todo{eight} leaf categories. The number in parentheses is the number of DBs exhibiting the corresponding symptom.

\textbf{Syntactic Error.} \cmmnt{\todo{176 (54.0\%)}~}\todob{226 (50.7\%)} of the~DBs~exhibit general syntactic errors that are similar to those in traditional~programs. It is the most common symptom.~Specifically,~\cmmnt{\todo{92 (28.2\%)}~}\todob{114~(25.6\%)}~of the~DBs manifest \textit{Element Not Found} errors; i.e., the used~syntactic elements like module, class, function, key and attribute cannot~be~retrieved. 
Further, \cmmnt{\todo{32 (9.8\%)}}\todob{36 (8.1\%)}~of~the~DBs~exhibit~\textit{Type Mismatch} errors; i.e., the variable type is inconsistent with the one that is expected. In addition, \cmmnt{\todo{21 (6.4\%)}~}\todob{25 (5.6\%)} and \cmmnt{\todo{16 (4.9\%)}~}\todob{18 (4.0\%)} of the DBs result~in~\textit{Illegal Value} and \textit{Illegal Argument} errors respectively, where a variable receives an illegal value, and a function call receives an illegal~argument. Moreover, \cmmnt{\todo{7 (2.1\%)}~}\todob{13 (2.9\%)} of the DBs report \textit{Undefined Variable}~errors, denoting that the variable is not defined or initialized. Besides, some infrequent errors (e.g., compilation errors) are included in the \textit{Others} category, which account for \cmmnt{\todo{8 (2.5\%)}~}\todob{20 (4.5\%)} of the DBs.




\textbf{DL Specific Error/Anomaly.} \cmmnt{\todo{118 (36.2\%)}~}\todob{173 (38.8\%)} of the DBs~exhibit DL specific errors or anomalies. It is the second~most~common~symptom, and is divided into five leaf categories. \textit{Software Error/Anomaly}~means errors or anomalies raised by software dependencies,~accounting for \cmmnt{\todo{64 (19.6\%)}~}\todob{108 (24.2\%)} of the DBs. There are four cases. (1) \cmmnt{\todo{14 (4.3\%)}~}\todob{18 (4.0\%)}~of~the~DBs~exhibit software~internal errors, indicated by an error message that contains the software name, e.g., CUDA\_ERROR\_UNKNOWN. 
(2) \cmmnt{\todo{32 (9.8\%)}~}\todob{59 (13.2\%)} of the DBs report that required software dependencies cannot be found. (3)~\cmmnt{\todo{11 (3.4\%)}~}\todob{11 (2.5\%)} of the DBs manifest dependency initialization failures, indicating that  required dependencies are not properly set up. (4) \cmmnt{\todo{7 (2.1\%)}~}\todob{20 (4.5\%)} of the DBs report~that~required software dependency~versions do not match.

Moreover, \textit{Hardware Error/Anomaly} denotes errors~or~anomalies raised by hardware dependencies; e.g., the GPU card~is~not correctly connected. It accounts for \cmmnt{\todo{10~(3.1\%)}~}\todob{14 (3.1\%)}~of~the DBs. Further,~\cmmnt{\todo{20~(6.1\%)}~}\todob{23~(5.2\%)}~of the DBs manifest \textit{Behavior Anomaly}, e.g., abnormal accuracy metrics and unexpected return values of APIs. In addition, \cmmnt{\todo{15 (4.6\%)}~}\todob{18 (4.0\%)} of the DBs exhibit~\textit{Model~Error}, which is indicated by an error message that contains~model elements, e.g., computation operator missing, model save/load failure, tensor conversion error, and layer unrecognized. Besides, \cmmnt{\todo{9 (2.8\%)}~}\todob{10 (2.2\%)} of the DBs manifest \textit{Data Anomaly}, reporting~that input data has abnormal values or mismatched property (e.g., size). 

\textbf{Performance Anomaly.} \cmmnt{\todo{17 (5.2\%)}~}\todob{21 (4.7\%)} of the DBs manifest~abnormal performance with respect to execution time, memory usage~and~processor usage. Specifically, \cmmnt{\todo{8 (2.5\%)}~}\todob{10 (2.2\%)} of the DBs exhibit \textit{Long Execution Time}; i.e., a program takes a long time to initialize or execute DL tasks, or even hangs in the middle of the execution. Further, \cmmnt{\todo{8 (2.5\%)}~}\todob{9 (2.0\%)} of the DBs cause \textit{Memory Anomaly}, including abnormal memory~utilization, memory leak, or even out of memory errors. Besides, \cmmnt{\todo{one}~}\todob{two} DBs result in \textit{Processor Anomaly} (i.e., high GPU utilization).

\textbf{Termination.} \cmmnt{\todo{11 (3.4\%)}~}\todob{20 (4.5\%)} of the DBs caused the program~directly terminated without any informative error code or error message. For example, it only reports a segmentation fault,~or~it simply reports that the task is killed or canceled.


\textbf{Warning.} \cmmnt{\todo{4 (1.2\%)}~}\todob{6 (1.3\%)} of the DBs show warning messages,~including warnings about function change, version compatibility, and~semantic mismatch in API arguments. For example,~a~version compatibility warning reveals that the installed version~violates the working version requirements. These warnings forecast the potential DBs due to using versions with changed elements.

 \todoq{\textbf{Comparison to DBs in Other Domains.} Compared to previous work, distinct symptoms of the DBs in our study are highlighted in dotted rectangles~in~Fig.~\ref{fig:error}, which include \textit{Hardware Error/Anomaly}, \textit{Model Error}, \textit{Data Anomaly} and \textit{Performance Anomaly}. They account for 63 (14.1\%) of the 446 DBs. 
These differences owe to the~fact that previous work is focused on DBs raised in homogeneous dependencies in the \textit{Application} and \textit{Library} layer in traditional software applications, while DBs across heterogeneous dependencies are not studied. Our study investigates DBs across the whole DL stack to collect symptoms revealed not only in dependencies within one layer but also in dependencies across layers.
}

\begin{tcolorbox}[size=title, opacityfill=0.15]
\textit{\textbf{Summary.}} General syntactic errors and DL specific errors and anomalies are the most common symptoms, which account~for \cmmnt{\todo{90.2\%}, 294/326, 399/446~}\todob{89.4\%} of the DBs and mostly cause crashes. Besides,~\cmmnt{\todo{5.2\%}~}\todob{4.7\%}~of~the DBs slow executions down~or~consume high resources. \tododel{These severe consequences of DBs motivate the significance of DBs.} \todoq{These wide-ranging impacts motivate the importance of DBs.} 

\end{tcolorbox}

\subsection{Exposing Stage and Dependency}\label{sec:exposing}

We identify the stage and dependency where the symptom~of each~DB is~exposed, and analyze DB distribution over them.

\begin{figure}[!t]
    \centering
    \includegraphics[width=0.45\textwidth]{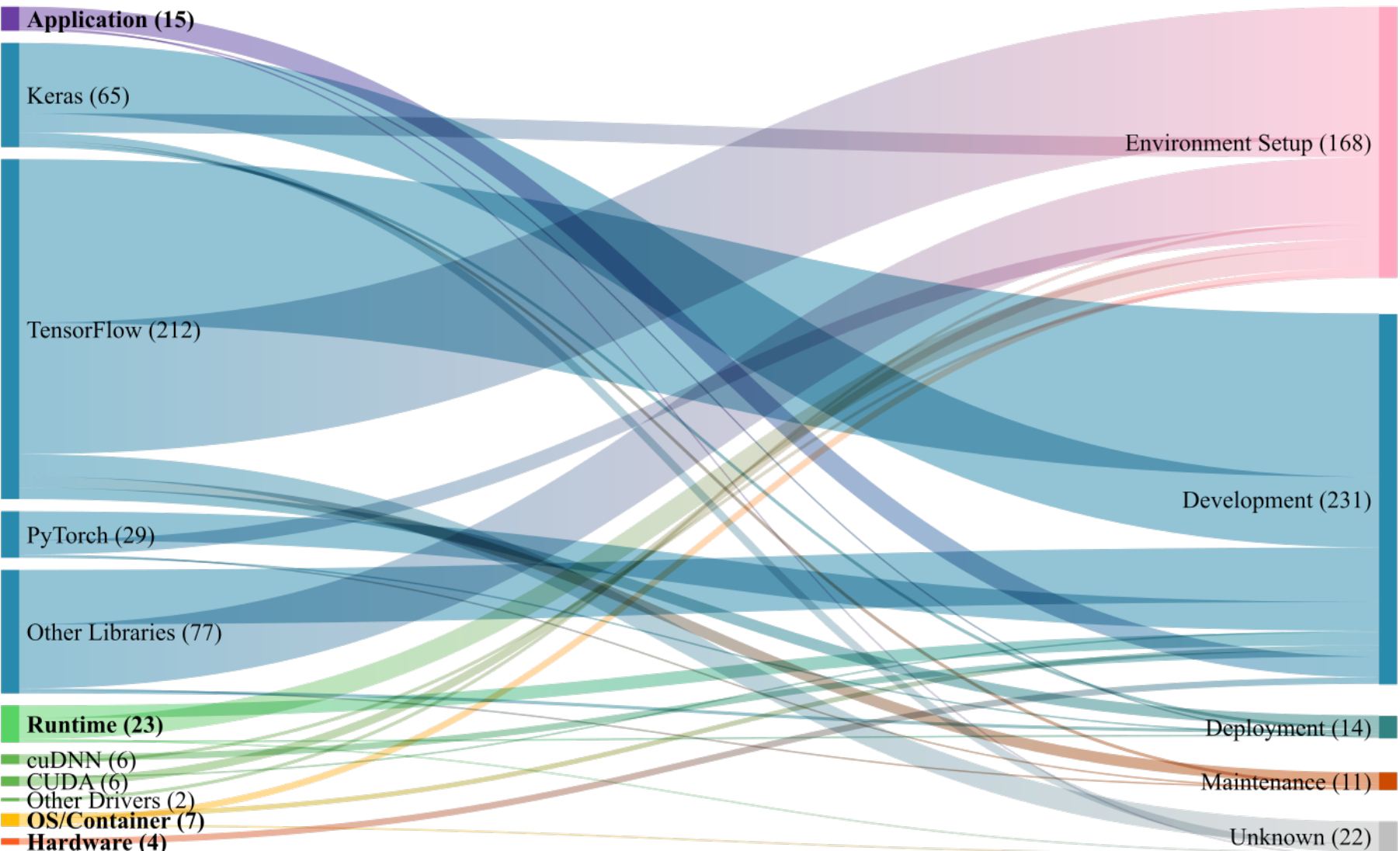}
    \vspace{-5pt}
    \caption{Exposing Dependency vs. Exposing Stage}
    \label{fig:exposing_library_stage}
\end{figure}

\textbf{Exposing Stage Analysis.} We classify the entire lifecycle of~engineering DL applications into four stages, i.e., environment setup,~development, deployment, and maintenance. 
We report the DB distribution over the exposing stages~in~the~right part of Fig.~\ref{fig:exposing_library_stage}. Development is the most bug-affecting stage, where~\cmmnt{\todo{200 (61.4\%)}~}\todob{231 (51.8\%)} of the DBs are exposed. This indicates~that although~the~setup process of DL stack is presumably finished, \tododel{a majority of DBs}\todoq{more than half of the DBs}~will~not occur until DL application~development. Environment setup is the second most bug-affecting stage, where \cmmnt{\todo{84 (25.8\%)}~}\todob{168 (37.7\%)} of the DBs are exposed. It indicates that the setup of a feasible DL stack is \tododel{a non-trivial task}\todoq{not easy}. 
\todob{Apart from the two dominating stages, deployment exposes 14 (3.1\%) and maintenance exposes~11~(2.5\%)~of~the~DBs,~which~are~relatively smaller than in environment setup and development.} 
The remaining \cmmnt{\todo{22 (6.7\%)}~}\todob{22 (4.9\%)} DBs have no clear indication about~the~exposing stage, and thus are included in the \textit{Unknown}~category.


\begin{tcolorbox}[size=title, opacityfill=0.15]
\textit{\textbf{Summary.}} The most bug-affecting stages are \tododel{environment setup and development}\todoq{development and environment setup}, exposing \cmmnt{\todo{87.1\%} 284/326, ~}\tododel{\todob{89.5\%}}\todoq{51.8\%} and \todoq{37.7\%} of the DBs.
\end{tcolorbox}

\textbf{Exposing Dependency Analysis.} We show the DB distribution over the exposing dependencies in the left part~of~Fig.~\ref{fig:exposing_library_stage}, which~is~organized by the layer hierarchy in DL stack (see~Sec. \ref{sec:stack}) with dominating dependencies separately highlighted. The \textit{Library}~layer~is~the~most bug-affecting layer, where \cmmnt{\todo{287 (88.0\%)}~}\todob{383 (85.9\%)} of the DBs are exposed. Specifically, Keras, TensorFlow and PyTorch in the \textit{Library} layer expose~\cmmnt{\todo{65 (19.9\%)}~}\todob{65 (14.6\%)}, \cmmnt{\todo{171 (52.5\%)}~}\todob{212 (47.5\%)}~and \cmmnt{\todo{16 (4.9\%)}~}\todob{29 (6.5\%)} of the DBs respectively, which are the most bug-affecting libraries. This is reasonable as they~are~currently the most popular DL frameworks. The \textit{Application} layer~exposes \cmmnt{\todo{12 (3.7\%)}~}\todob{15 (3.4\%)} of the DBs, while the \textit{Driver}~layer~exposes~\cmmnt{\todo{11~(3.4\%)}~}\todob{14~(3.1\%)} of the DBs. CUDA and cuDNN \tododel{are the most bug-affecting drivers,~and} 
\todob{both expose 6 (1.3\%) of the DBs}. Besides, there are at most \cmmnt{\todo{7 (2.1\%)}~}\todob{23 (5.2\%)} of the DBs that are exposed at the dependencies at the \textit{Runtime}, \textit{OS/Container} or \textit{Hardware} layer.



The Sankey diagram in Fig.~\ref{fig:exposing_library_stage} illustrates where the DBs exposed in a dependency are exposed across the lifecycle stages.~The width of the flow is proportional to the number of DBs. Generally,~a DB can be exposed at any dependency at any layer in DL stack~at~any stage of the~engineering lifecycle. This indicates the complexity~of~DBs.
\vspace{-2pt}
\begin{tcolorbox}[size=title, opacityfill=0.15]
\textit{\textbf{Summary.}} \textit{Library}, \textit{Application} and \textit{Driver} are the~most bug-affecting stack layers. 
Keras, TensorFlow and PyTorch are the most bug-affecting~libraries\tododel{, while CUDA and cuDNN are the most bug-affecting drivers}. 
\end{tcolorbox}


\section{RQ2: Root Cause Analysis}\label{sec:root-cause}

We report the taxonomy of DB root causes, and analyze the stages and dependencies where root causes are introduced.

\begin{figure}[!t]
    \centering
    \includegraphics[width=0.42\textwidth]{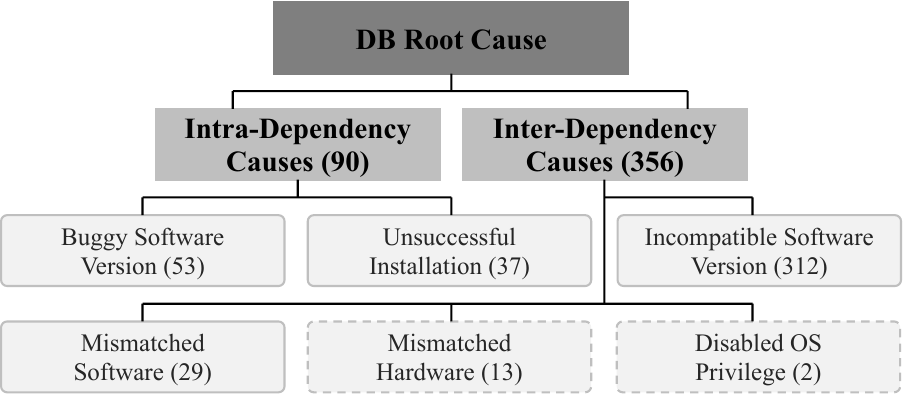}
    \vspace{-7pt}
    \caption{\todoq{Taxonomy of DB Root Causes}}
    \label{fig:root_cause}
\end{figure}

\subsection{Root Cause Taxonomy}\label{sec:root_cause}

The taxonomy of DB root causes is shown in Fig.~\ref{fig:root_cause}.~We~first classify the root causes based on the criterion that whether~a~DB is~caused by one dependency (i.e., \textit{Intra-Dependency Cause})~or by constraints among dependencies across DL stack (i.e., \textit{Inter-Dependency Cause}). Then, we summarize six leaf categories.

\textbf{Intra-Dependency Cause.} \cmmnt{\todo{48 (14.7\%)}~}\todob{90 (20.2\%)} of the DBs~are~caused~solely by one dependency itself, and are divided into two leaf categories. Particularly, \cmmnt{\todo{40 (12.3\%)}~}\todob{53 (11.9\%)} of the DBs are caused by \textit{Buggy Software~Version}; i.e., a DB is caused by triggering~bugs in~software dependencies in DL stack. 
Moreover, \cmmnt{\todo{8 (2.5\%)}~}\todob{37 (8.3\%)} of the DBs are caused~by~\textit{Unsuccessful Installation} of dependencies. There are two cases.~(1)~Dependency~installation does~not complete. For example, a developer found that~there~was~no file named \texttt{cudnn64\_6.dll}, which~was~caused by the missed installation of cuDNN on his/her machine~\cite{cudnn6}.~(2)~Dependency installation completes, but lacks proper path configuration (e.g., missing path configuration or configuring incorrect path). 




\textbf{Inter-Dependency Cause.} \cmmnt{\todo{278 (85.3\%)}~}\todob{356 (79.8\%)} of the DBs are caused by constraints among software and hardware dependencies across DL stack; i.e., multiple dependencies have to be considered together~to have a feasible DL stack, otherwise, DBs might be introduced.~It~is divided into four leaf categories. Specifically, \cmmnt{\todo{248 (76.1\%)}~}\todob{312 (70.0\%)}~of the~DBs are caused~by~\textit{Incompatible Software Version},~which is the most~common~root~cause. An incompatible software version is introduced~if it violates the version constraint that has to be satisfied for~it~to~work~with other dependencies. For \cmmnt{\todo{56}~}\todob{58} of the \cmmnt{\todo{248}~}\todob{312} DBs, detailed API-level incompatibility information is provided in the post/issue, and we classify the incompatibility based on API changes~\cite{brito2018apidiff}, i.e., API removal, API addition, API replacement, API movement, API parameter list change, API~renaming, and API behavior change. API addition, API behavior change and API removal are the most~common root causes of API incompatibility, which respectively account for at least \cmmnt{\todo{19 (5.8\%)}, \todo{13 (4.0\%)} and \todo{10 (3.1\%)}~}\todob{20 (4.5\%), 13 (2.9\%)~and~11 (2.5\%)} of the DBs. API replacement, API parameter list~change, API movement and API renaming respectively cause at least \cmmnt{\todo{4, 4, 3 and 3}~}\todob{4, 4, 3 and 3} DBs. For the \cmmnt{\todo{138}~}\todob{177} of the \cmmnt{\todo{248}~}\todob{311} DBs,~we~can~only distinguish whether they are caused by backward incompatibility (for \cmmnt{\todo{80 (24.5\%)}~}\todob{101 (22.6\%)} of the DBs) or forward incompatibility (for \cmmnt{\todo{58 (17.8\%)}~}\todob{76 (17.0\%)} of the DBs). For the remaining \cmmnt{\todo{54}~}\todob{77} of the \cmmnt{\todo{248}~}\todob{311} DBs, we can only determine they are caused by incompatibility due to the limited information in the posts/issues.




\cmmnt{\todo{19 (5.8\%)}}~\todob{29 (6.5\%)} of the DBs are caused by \textit{Mismatched Software}; i.e.,~while different software can provide similar functionalities, only some of them can work with the other dependencies~in~DL stack, but others are regarded as mismatched.~Specifically,~\cmmnt{\todo{11}~}\todob{15} of the~\cmmnt{\todo{19}~}DBs~are~caused by selecting wrong software as dependency. For example, the DL~framework Keras and the \texttt{tf.keras} module introduced in TensorFlow~1.10 provide similar APIs, but Keras does not support TensorFlow~2.0.~In that~sense,~if TensorFlow 2.0 is used in DL stack, Keras would~be~mismatched and thus cannot be used~\cite{kerasnotcompatible}. Further,~\cmmnt{\todo{5}~}\todob{11} of the~\cmmnt{\todo{19}~}DBs~are~caused by choosing wrong software distribution. For example,~the~official pre-built TensorFlow 2.0 requires CUDA Toolkit 10.0. Developers have to re-build TensorFlow 2.0 with CUDA Toolkit 10.1 to work with CUDA Toolkit 10.1. Thus, using~pre-built TensorFlow 2.0 with CUDA Toolkit 10.1 could~cause~a~DB~\cite{prebuiltwontwork}. Further, \cmmnt{\todo{3}~}\todob{3} of the \cmmnt{\todo{19}~}DBs~are caused~by~selecting~multiple~conflicting software. For example,~loading both TensorFlow and TensorFlow-gpu \cite{conflictgpu} 
would cause a DB.




\cmmnt{\todo{9 (2.8\%)}~}\todob{13 (2.9\%)} of the DBs are caused by \textit{Mismatched Hardware}; i.e., the hardware does not meet requirements of dependencies in upper stack layers. For example, TensorFlow~1.6~used~AVX feature of CPUs, which~is~supported by Sandy Bridge or newer CPU architectures. Hence, using TensorFlow with non-AVX~CPUs~would~cause~a~DB~\cite{avx}.

\cmmnt{\todo{2 (0.6\%)}}~\todob{2 (0.4\%)} of the DBs are caused by \textit{Disabled OS Privilege}; i.e.,~permissions required by software dependencies are not allowed from the OS or container. For example, System Integrity Protection (SIP) is enabled on MacOS~10.11~to~prevent~unauthorized code execution, but SIP prevents~a~path~variable from being overridden, causing dependencies not found~\cite{macsip}.

\todoq{\textbf{Comparison to DBs in Other Domains.} Compared to previous work, distinct root causes of the DBs in our study are highlighted in dotted rectangles in Fig.~\ref{fig:root_cause}, which include \textit{Mismatched Hardware} and \textit{Disabled OS Privilege}. They account for 15 (3.4\%) of the 446~DBs. It is worth mentioning that although most root causes are shared with previous work, the dependencies that cause DBs can be different (see Sec.~\ref{sec:intr}) as our study further considers the \textit{Runtime}, \textit{Driver}, \textit{OS/Container} and \textit{Hardware} layers.
}

\begin{tcolorbox}[size=title, opacityfill=0.15]
\textit{\textbf{Summary.}} Violation of constraints among dependencies~in DL stack causes \cmmnt{\todo{85.3\%}~}\todob{79.8\%} of the DBs, where incompatible~software version is the \tododel{dominating}\todoq{major} root~cause. Moreover, bugs~in software dependencies cause \cmmnt{\todo{12.3\%}~}\todob{11.9\%}~of~the~DBs.
\end{tcolorbox}

\begin{figure}[!t]
    \centering
    \includegraphics[width=0.44\textwidth]{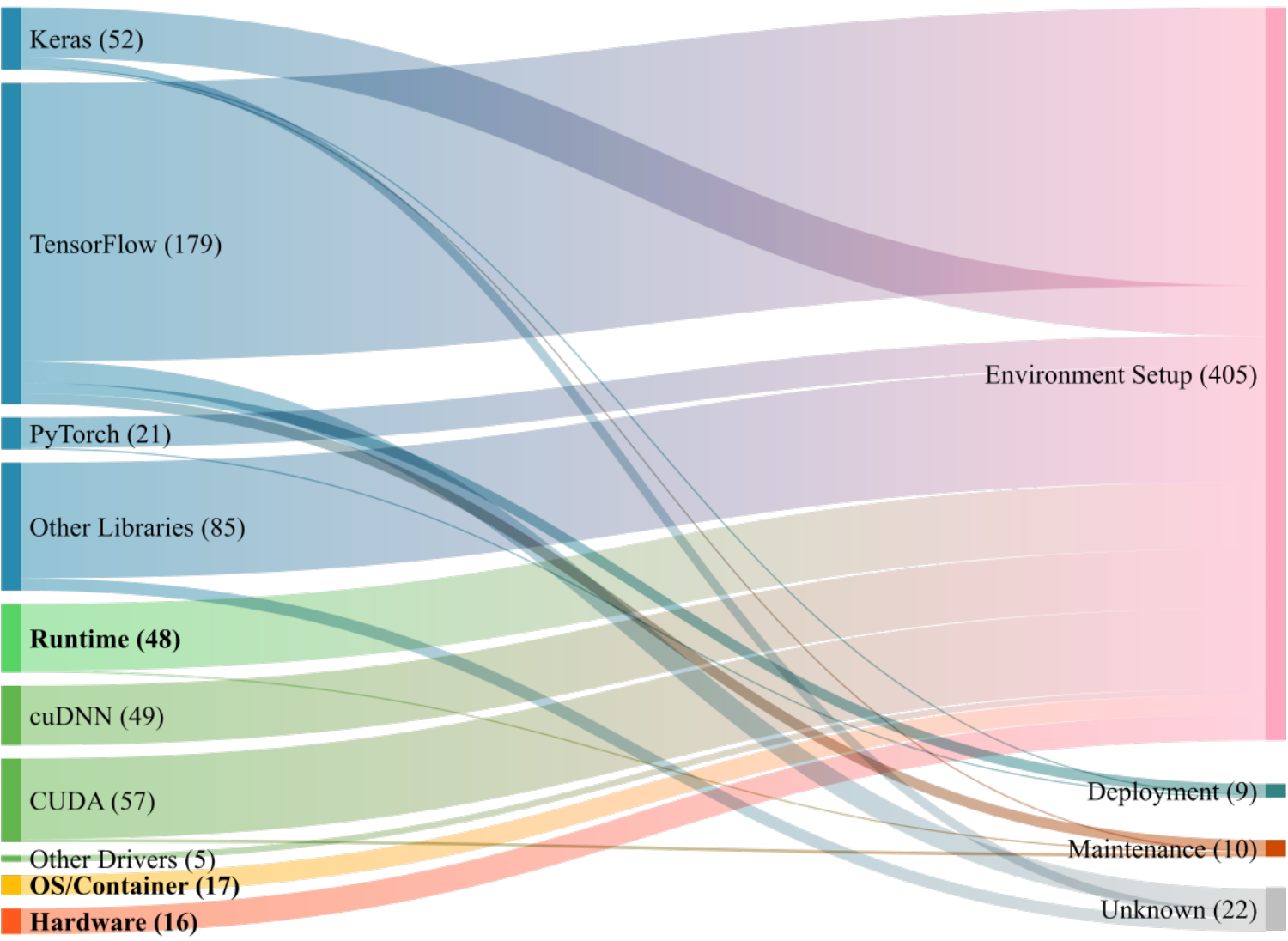}
    \vspace{-5pt}
    \caption{Introducing Dependency vs. Introducing Stage}
    \label{fig:root_cause_stages}
\end{figure}

\subsection{Introducing Stage and Dependency}\label{sec:intr}

We locate the stage and dependency where the root cause~of each~DB is introduced, and analyze DB distribution over them.

\textbf{Introducing Stage Analysis.} The taxonomy of stages~is~the same to the one in Sec.~\ref{sec:exposing}. We show the DB distribution~over~the introducing stages~in~the~right part of Fig.~\ref{fig:root_cause_stages}. Environment setup~is~the most bug-prone stage, where~\cmmnt{\todo{285 (87.4\%)}~}\todob{405 (90.8\%)}~of~the DBs are introduced, while no DB is introduced in development because the DL stack is already determined in environment setup. It indicates that the setup~of~a~feasible DL stack is important but challenging. Besides, deployment and maintenance introduce \cmmnt{\todo{9 (2.8\%)}~}\todob{9 (2.0\%)} and \cmmnt{\todo{10 (3.0\%)}~}\todob{10 (3.2\%)} of the DBs. The remaining \cmmnt{\todo{22 (6.7\%)}~}\todob{22 (4.9\%)} DBs have no clear indication~about the~introducing stage, and thus are put in the \textit{Unknown}~category.

\begin{tcolorbox}[size=title, opacityfill=0.15]
\textit{\textbf{Summary.}} The most bug-prone stage is environment setup, which introduces \cmmnt{\todo{87.4\%}~}\todob{90.8\%} of the DBs.
\end{tcolorbox}

\begin{figure*}[!t]
    \centering
    \includegraphics[width=0.83\textwidth]{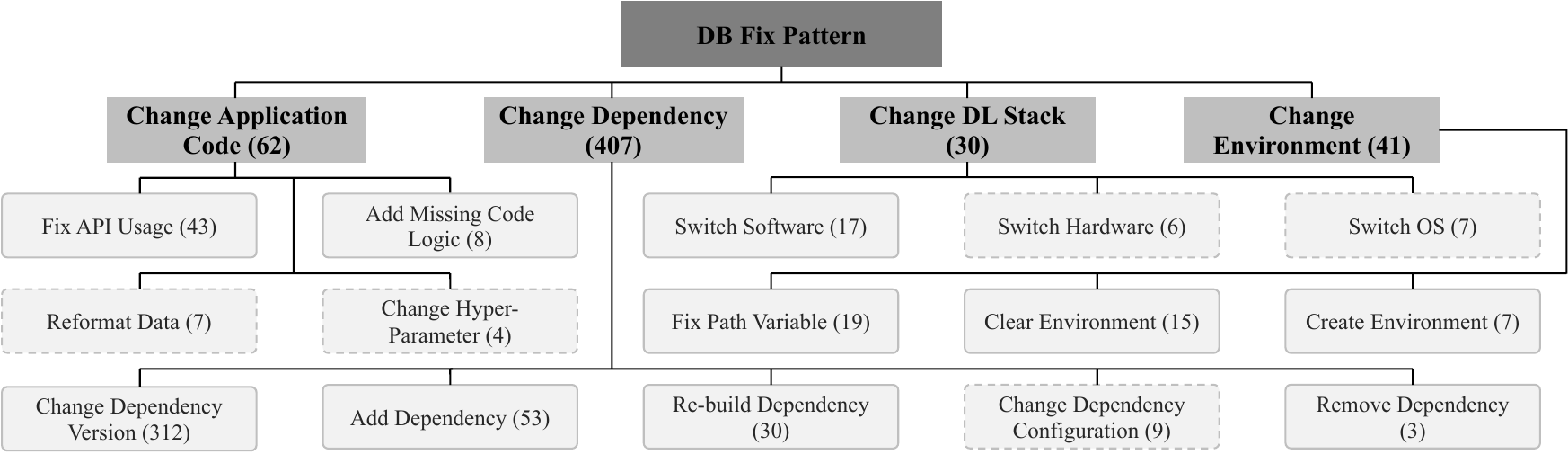}
    \vspace{-5pt}
    \caption{\todoq{Taxonomy of DB Fix Patterns}}
    \label{fig:fix_taxonomy}
\end{figure*}

\textbf{Introducing Dependency Analysis.} For the DBs caused~by inter-dependency causes, their root causes can be introduced by~any of the involved dependencies. For example, a DB is caused~by~version~constraint violation between TensorFlow~and CUDA, and then~both~TensorFlow and CUDA~can~be~the~introducing dependency of this DB. If there is no clear indication about the introducing dependency~in~the posts/issues,~we~consider all involved dependencies as the introducing~dependencies; otherwise, we use the introducing dependency that~is decided in the posts/issues. \todohs{It is worth mentioning that the introducing dependency of \todo{114} DBs~are only mentioned in the answers of the posts/issues, indicating that questioners are not aware of the introducing~dependency. In \todo{258} of the DBs, questioners only mention the dependency list to provide more detail, but there is no clue indicating that they are aware~of~the introducing dependency. In \todo{74} of the DBs, questioners indicate assumptions on the introducing dependency.} Specifically, of the \cmmnt{\todo{278}~}\todob{356} DBs that are caused by inter-dependency causes, \cmmnt{\todo{257}~}\todob{336} DBs have their introducing dependencies clearly indicated in the posts/issues.

We show the DB distribution over the introducing dependencies in the left part of Fig.~\ref{fig:root_cause_stages}, which~is~organized~in~the~same way~in Fig.~\ref{fig:exposing_library_stage}. No DB is introduced~in the \textit{Application}~layer as~it~is~the client of dependencies. The~\textit{Library} and \textit{Driver}~layers are the~most bug-prone layers, respectively introducing \cmmnt{\todo{246 (75.5\%)}~}\todob{301 (67.5\%)} and \cmmnt{\todo{59 (18.1\%)}~}\todob{94 (21.1\%)}~of~the~DBs. The number is larger~than the summation of DBs in all~libraries or drivers because~a~DB can have multiple introducing dependencies. \tododel{Specifically,~Keras, TensorFlow and PyTorch in the \textit{Library} layer~are the most bug-prone libraries, introducing~\cmmnt{\todo{51 (15.6\%)}~}\todob{52 (11.7\%)}, \cmmnt{\todo{167 (51.2\%)}~}\todob{179 (40.1\%)} and \cmmnt{\todo{12 (3.7\%)}~}\todob{21 (4.7\%)} of the DBs. CUDA and cuDNN are the most bug-prone drivers, introducing \cmmnt{\todo{34 (10.4\%)}~}\todob{57 (12.8\%) and 49 (11.0\%)} of the DBs.}\todoq{Specifically,~Keras, TensorFlow and PyTorch introduce the most bugs in the \textit{Library} layer, introducing~\cmmnt{\todo{51 (15.6\%)}~}\todob{52 (11.7\%)}, \cmmnt{\todo{167 (51.2\%)}~}\todob{179 (40.1\%)} and \cmmnt{\todo{12 (3.7\%)}~}\todob{21 (4.7\%)} of the DBs. CUDA and cuDNN introduce the most bugs in the \textit{Driver} layer, introducing \cmmnt{\todo{34 (10.4\%)}~}\todob{57 (12.8\%) and 49 (11.0\%)} of the DBs.} There~are at most \cmmnt{\todo{22 (6.7\%)}~}\todob{48 (10.8)} of the DBs introduced at the dependencies~at the \textit{Runtime}, \textit{OS/Container} or \textit{Hardware}~layer.

Besides, the Sankey diagram in Fig.~\ref{fig:root_cause_stages} shows where~the DBs introduced in a dependency are introduced across the~lifecycle stages. Generally,~a DB can be introduced at any dependency at any layer (except for \textit{Application}) at~any lifecycle stage (except for development). It reveals the complexity~of~DB localization.

\begin{tcolorbox}[size=title, opacityfill=0.15]
\textit{\textbf{Summary.}} \textit{Library} and \textit{Driver} are the most bug-prone stack layers, which introduce \cmmnt{\todo{90.5\%}~382}\tododel{\todob{85.7\%}}\todoq{301 (67.5\%) and 94 (21.1\%)} of the DBs. \tododel{Keras, TensorFlow and PyTorch are the most bug-prone libraries, while CUDA and cuDNN are the most bug-prone drivers.} \todoq{Keras, TensorFlow and PyTorch introduce the most bugs in the \textit{Library} layer, while CUDA and cuDNN introduce the most bugs in the \textit{Driver} layer.} 
\end{tcolorbox}

\subsection{Introducing and Exposing Dependency}


We further analyze where the DBs introduced in a dependency~are~exposed across the dependencies in DL stack. 
Overall, \cmmnt{\todo{142 (43.6\%)}~}\todob{227 (50.9\%)}~of~the DBs are not introduced~and~exposed~in~the same dependency. For example, \cmmnt{\todo{32 (9.8\%)}~}\todob{33 (7.4\%)} of the DBs introduced in TensorFlow are exposed in Keras, and \cmmnt{\todo{41 (12.6\%)}~}\todob{60 (13.5\%)} of the DBs introduced in CUDA~and~cuDNN~are~exposed~in TensorFlow. At the stack layer level, \cmmnt{\todo{97 (29.8\%)}~}\todob{162 (36.3\%)} of the~DBs are not introduced and exposed at the same stack layer.~For~example, \cmmnt{\todo{8}~}\todob{12 (2.7\%)} of the DBs introduced at the \textit{Hardware} layer are exposed~at~the~\todo{\textit{Library}} layer. These results indicate that~DB localization need systematic knowledge of the entire DL~stack.


\begin{tcolorbox}[size=title, opacityfill=0.15]
\textit{\textbf{Summary.}} \cmmnt{\todo{142 (43.6\%)}~}\todob{227 (50.9\%)} of the DBs are not introduced~and exposed in the same dependency, and \cmmnt{\todo{97 (29.8\%)}~}\todob{162 (36.3\%)}~of~the DBs are not introduced and exposed at the same layer.
\end{tcolorbox}

%
%


\section{RQ3: Fix Pattern Analysis}\label{sec:fix-pattern}

We present the taxonomy of DB fix patterns, and report~their distribution for root causes and the knowledge source~of fixing.

\subsection{Fix Pattern Taxonomy}

The taxonomy of DB fix patterns is listed in Fig.~\ref{fig:fix_taxonomy}. It is grouped~into four inner categories (i.e., \textit{Change Application Code}, \textit{Change Dependency}, \textit{Change DL Stack} and \textit{Change Environment})~and~15~leaf categories. A DB can be fixed by applying multiple fix patterns.~Hence, the summation of the number of DBs in Fig.~\ref{fig:fix_taxonomy} is larger than \cmmnt{\todo{326}~}\todob{446}.


\textbf{Change Application Code.} \cmmnt{\todo{59 (18.1\%)}~}\todob{62 (13.9\%)} of the DBs~are~fixed via changing the application code although their root causes are not introduced by the application. Specifically, \textit{Fixing API Usage}~is~used~to fix \cmmnt{\todo{41 (12.6\%)}~}\todob{43 (9.6\%)} of the DBs; i.e.,~the~library~API usage~has~to~be changed with the incompatible library version evolution. Moreover,~\textit{Adding Missing Code Logic} is utilized~to fix \cmmnt{\todo{7 (2.1\%)}~}\todob{8 (1.8\%)} of the DBs.~In~such~cases, some library~APIs~are removed or the behavior of some library APIs is changed,~and hence developers have to implement the code logic of these library APIs by themselves at the application code~level.~Further, \textit{Reformatting Data} is used to fix \cmmnt{\todo{7 (2.1\%)}~}\todob{7 (1.6\%)} of the DBs for making the data format compatible with~the~changed~library~APIs. Besides, \textit{Changing Hyper-Parameter} (e.g., batch size~and~learning rate) is used to fix \cmmnt{\todo{4 (1.2\%)}~}\todob{4 (0.9\%)} of the DBs, because the constraints on hyper-parameters are changed with library version evolution.



\textbf{Change Dependency.} This is the most common fix pattern, which is leveraged to fix \cmmnt{\todo{304 (93.3\%)}~}\todob{407 (91.3\%)} of the DBs. In particular, \textit{Changing Dependency Version} is used~to~fix \cmmnt{\todo{256 (78.5\%)}~}\todob{312 (70.0\%)}~of~the DBs, indicating that it is the most common pattern to fix DBs. Of these \cmmnt{\todo{256}~}\todob{312} DBs, upgrading dependency version~is~used~in~the fix of \cmmnt{\todo{153}~}\todob{188} DBs, and downgrading dependency version~is~used~in the fix of \cmmnt{\todo{104}~}\todob{122} DBs. \todo{In \cmmnt{17~}\todob{22} of the DBs, dependency version~is changed but there is no clear indication in the posts/issues to determine upgrade or downgrade.} Further, \textit{Adding Dependency} is used to fix \cmmnt{\todo{23 (7.1\%)}~}\todob{53 (11.9\%)} of the DBs where~some~required~dependencies are missing or not successfully installed. Moreover, \textit{Re-building Dependency}~is used to fix \cmmnt{\todo{19 (5.8\%)}~}\todob{30 (6.7\%)}~of~the~DBs.~In such cases, the source code~of~dependencies is re-built~with other required dependencies to properly work with them,~or~the source code of dependencies is first changed (e.g., to fix~bugs~or to~remove incompatibilities) and then re-built, potentially~because~of~the~huge maintenance effort~in~changing~dependency versions. In addition, \textit{Changing Dependency Configuration} is leveraged to fix \cmmnt{\todo{4 (1.2\%)}~}\todob{9 (2.0\%)}~of~the DBs, e.g., disabling SIP in MacOS. Besides, \textit{Removing Dependency} is applied to fix \cmmnt{\todo{2 (0.6\%)}}\todob{3 (0.7\%)} of the DBs in order to remove conflicted dependencies.


\textbf{Change DL Stack.} \cmmnt{\todo{22 (6.7\%)}~}\todob{30 (6.7\%)} of the DBs are fixed~by~changing the DL stack; i.e., some dependencies are switched~to~alternatives, and the DL stack becomes fundamentally different. It is divided into three leaf categories,~i.e.,~\textit{Switching Software} (libraries, drivers and runtimes), \textit{Switching Hardware} and \textit{Switching OS}, accounting for \cmmnt{\todo{11~(3.4\%)}~}\todob{17~(3.8\%)}, \cmmnt{\todo{6 (1.3\%)}~}\todob{6 (1.8\%)} and \cmmnt{\todo{5 (1.1\%)}~}\todob{7 (1.6\%)} of the DBs.~For~example, a developer switched OS to Ubuntu to support distributed TensorFlow~\cite{switchos}.


\textbf{Change Environment.} \cmmnt{\todo{24 (7.4\%)}~}\todob{41 (9.2\%)} of the DBs are fixed by changing the environment where libraries, drivers and runtimes can be found. Specifically, \textit{Fixing Path Variable} is used to fix \cmmnt{\todo{9 (2.8\%})~}\todob{19 (4.3\%)} of the DBs; i.e., the path variable is fixed to point to the correct directory that contains the required dependencies. Besides, \textit{Clearing~Environment} and \textit{Creating Environment} are used to respectively fix \cmmnt{\todo{9 (2.8\%})~}\todob{15 (3.4\%)} and \cmmnt{\todo{6 (1.8\%})~}\todob{7 (1.6\%)} of the DBs. In these cases, the virtual environment (i.e., a directory that contains a specific collection of installed packages) of package managers (e.g., pip and conda) is cleared or created.


Notice that \cmmnt{\todo{279 (85.6\%)}~}\todob{389 (87.2\%)} of the DBs can be fixed by applying one fix pattern, while \cmmnt{\todo{64 (19.6\%)}~}\todob{73 (16.4\%)}, \cmmnt{\todo{18 (5.5\%)}~}\todob{20 (4.5\%)} and \cmmnt{\todo{3 (0.9\%)}~}\todob{3 (0.7\%)}~of~the DBs can be fixed by combining two, three and four fix patterns at the same time. The summation here is larger than \cmmnt{\todo{326}~}\todob{446}~as~\cmmnt{\todob{\todo{36}~}}\todob{37} DBs can be fixed by different combinations of fix patterns. 


\todoq{\textbf{Comparison to DBs in Other Domains.} Compared to previous work, distinct fix patterns of the DBs in our study are highlighted in dotted rectangles in Fig.~\ref{fig:fix_taxonomy}, which include \textit{Reformat Data}, \textit{Change Hyper-Parameter}, \textit{Switch Hardware}, \textit{Switch OS}, and \textit{Change Dependency Configuration}. They are used to fix 32 (7.2\%) of the 446 DBs. 
Moreover, multiple fix patterns need to be combined to fix some DBs in DL stack, which is not the case in fixing dependency conflicts~\cite{artho2012software, patra2018conflictjs,wang2018dependency, huang2020interactive, wang2019could, wang2021will, wang2020watchman, wang2021hero} where only one fix pattern~is~needed.}

\begin{tcolorbox}[size=title, opacityfill=0.15]
\textit{\textbf{Summary.}} The most common fix pattern is to change~dependency versions, which is used to fix \cmmnt{\todo{78.5\%}~}\todob{70.0\%} of the DBs. 
\todob{Adding dependency} is the second~most~common pattern, which is leveraged to fix \cmmnt{\todo{12.6\%}~}\todob{11.9\%} of the DBs. \cmmnt{\todo{26.1\%}~}\todob{21.5\%} of the DBs can be fixed by combining multiple fix patterns.
\end{tcolorbox}

\subsection{Distribution of Fix Patterns for Root Causes}

%
%

\begin{figure}[!t]
    \centering
    \includegraphics[width=0.44\textwidth]{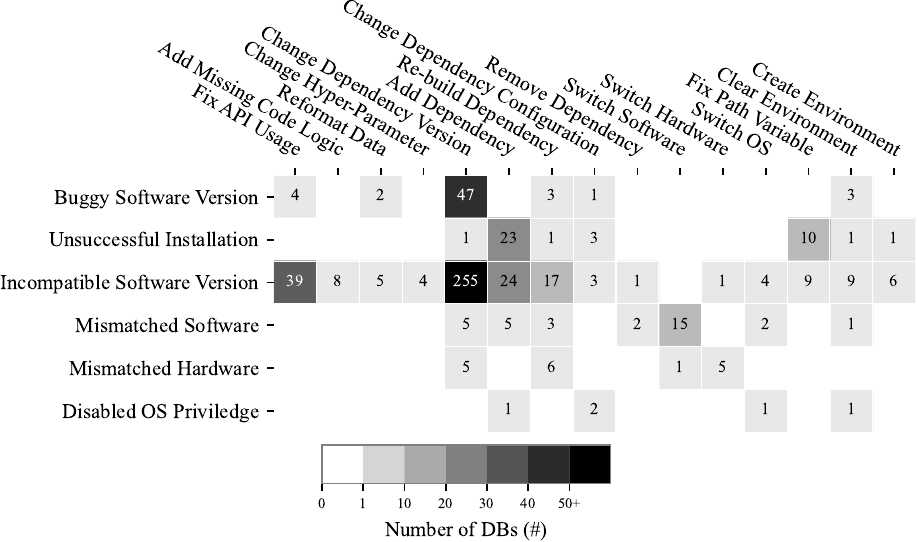}
    \vspace{-5pt}
    \caption{Distribution of Fix Patterns for Root Causes}
    \label{fig:heatmap_root_cause_fix}
\end{figure}

We report the distribution of fix patterns for root causes in Fig.~\ref{fig:heatmap_root_cause_fix}, where each cell denotes the number of DBs~that~are caused by~a~particular root cause and fixed by a particular~fix pattern. Specifically, except~for~\textit{Switching Software},~all~fix~patterns are utilized in fixing DBs that are caused by \textit{Incompatible~Software Version} for at least once. 
While \textit{Fixing API Usage} and \textit{Changing Dependency Version} are the two \tododel{dominating}\todoq{major}~fix patterns, there exist diverse ways~to~fix~the most common root cause~\textit{Incompatible~Software Version}. The challenge is to decide which fix pattern to use given a DB~context.



Further, \textit{Changing Dependency Version} is used in mitigating five root causes, and is involved in the fix for \cmmnt{\todo{256 (78.6\%})~}\todob{312 (70.0\%)}~of the~DBs. It has a strong correlation to the root causes \textit{Buggy Software Version} and \textit{Incompatible Software Version}. While the fix pattern itself~is~very simple, the key challenge~is~to~determine which dependency version to use for addressing~a~DB. Besides, \textit{Adding Dependency}, \textit{Re-building Dependency} and \textit{Clearing Environment} are the other three fix patterns spanning~at~least four root causes. Notice that \textit{Adding Dependency}~is~the~accompanied fix pattern for fixing DBs caused~by \textit{Incompatible Software Version}. For example, upgrading dependency version solves an \textit{Incompatible Software Version},~but~this~upgraded dependency version may further depend on a new dependency.

\begin{tcolorbox}[size=title, opacityfill=0.15]
\textit{\textbf{Summary.}} \textit{Incompatible Software Version} can be fixed~by diverse patterns, whereas \textit{Fixing API~Usage} and \textit{Changing Dependency Version} are the two \tododel{dominating}\todoq{major}~fix patterns. \textit{Changing Dependency Version} is also the \tododel{dominating}\todoq{major} fix pattern for \textit{Buggy Software Version}.
\end{tcolorbox}

\subsection{Knowledge Source of DB Fixing}

To fix a DB, developers usually rely on knowledge about~DL~stack, e.g., dependency version constraints~and~dependency bugs.~To characterize how fixes of DBs are derived, we investigate the knowledge sources that are used to fix DBs. We identify five knowledge sources. Multiple knowledge sources can be used in fixing one DB, and hence the summation of the number of DBs below is larger than \cmmnt{\todo{326}~}\todob{446}.


\textbf{Library Source Code.} \cmmnt{\todo{51 (15.6\%)}~}\todob{52 (11.7\%)} of the DBs~are fixed~after~digging into~the source code of libraries. The source code~of libraries~is a good~knowledge source to know library version evolution, e.g.,~how~a library API is renamed, and how a library~API's code logic~is~changed.

\textbf{Dependency Documentation.} \cmmnt{\todo{62 (19.0\%)}~}\todob{76 (17.0\%)} of the DBs~are fixed after looking into dependency documentation. Documentation of libraries, drivers and hardware often provide~informative~knowledge about dependency's~installation requirements and version~constraints.~For example, 
TensorFlow documentation~lists~both the hardware requirements and software~requirements \cite{tfinstall}.


\textbf{Issue Tracker.} \cmmnt{\todo{18 (5.5\%)}~}\todob{23 (5.2\%)}  of the DBs are fixed after being aware~of the dependency bugs. Such bugs are tracked on issue trackers with their symptoms and affected versions described.

\textbf{Other Online Resource}. \cmmnt{\todo{19 (5.8\%)}}\todob{22 (4.9\%)} of the DBs are fixed~by referencing other online resources, e.g., mailing lists, StackOverflow posts and technical blogs.

\textbf{Unknown.} For \cmmnt{\todo{200 (61.3\%)}~}\todob{309 (69.3\%)} of the DBs, there~is~no~clear indication about the used knowledge source in the posts/issues, and~thus we include them into the \textit{Unknown} category.~However, such posts/issues themselves become a knowledge source.


\begin{tcolorbox}[size=title, opacityfill=0.15]
\textit{\textbf{Summary.}} \todop{Library source code, dependency documentation, issue tracker, and other online resource are \todoq{important} knowledge sources that are directly leveraged to fix DBs.}
\end{tcolorbox}


\section{Implication,~Application~and~Threat}\label{sec:implication}

We discuss the implications of our study, demonstrate an application, and analyze the threats to our study.

\subsection{Implication to Developers and Researchers}


\textbf{\todoq{Application} Developers.} Our study uncovers~the~common DB symptoms~that developers should be aware of when engineering DL applications for detecting potential DBs as early as possible. Our study also~identifies the~common root causes and~fix~patterns of DBs that could~be~useful for \todoq{application} developers~to~diagnose, localize and fix DBs. Our study~also~shows the most \tododel{bug-prone}\todoq{bug-introducing} and bug-affecting dependencies where \todoq{application} developers should pay more attention when installing, using or maintaining them so that most DBs could be avoided or detected~at~the~first~place. Moreover, our findings provide some engineering suggestions. Application developers should be trained to have a comprehensive~understanding of the DL stack, as our study reports that~a~DB~could be introduced or exposed across the entire DL stack~and~engineering lifecycle.~In~this way, \todoq{application} developers are equipped with the sufficient knowledge to deal with DBs. Appilcation developers should carefully look into dependency documentation to learn version constraints,~and~be~aware~of~the~bugs and~API~changes~in~library version evolution. In this way, DBs~caused by the most common root causes (i.e., \textit{Buggy Software Version} and \textit{Incompatible Software Version}) might be effectively reduced.

\todoq{\textbf{Library Developers.} Our study reveals that around half of the DBs are not introduced and exposed in the same dependency.~This requires library developers to write informative error messages~in~exposing dependencies to help indicate the root causes in introducing dependencies. 
In addition, our study identifies \textit{Incompatible Software~Version} as the common root cause and \textit{Change~Dependency Version} as the common fix pattern for DBs. This highlights~the~importance of providing precise version constraints by library developers to allow application developers to follow and thus prevent DB occurrences. Furthermore, if library developers integrate certain version constraint checking in dependency installation scripts and provide potential version constraint violation hints for application developers, it would eliminate DBs at the first place. 
}

\textbf{Researchers.} Our findings provide future research implications in four directions. First, \textit{a dependency knowledge graph for the entire DL stack is needed to provide fundamental~knowledge for~the~ease~of dependency management.} As uncovered by our root cause analysis,~a diversity~of~dependency knowledge~is involved in DBs, e.g., version constraints among software~and hardware dependencies, bugs in dependencies, and API changes in version evolution. However, such knowledge is scattered across different sources,~e.g.,~documentation, issue tracker and source code, as revealed by our investigation~of~the knowledge source~of DB fixing. Online~resources like StackOverflow posts \todob{and GitHub issues} also provide practical solutions~to~fix~DBs. Hence, the main challenges to construct the knowledge graph are that i)~designing~a high-level schema to fuse various~knowledge into a graph, ii)~leveraging various techniques like natural language processing and program analysis to automatically extract knowledge~from~different sources and keep them up-to-date; and iii) developing graph analysis techniques~for~various dependency management tasks.~This knowledge graph serves~as the foundation of the following three research directions.~\todo{Along this direction, Ye~et~al. \cite{Ye2022} and Cheng et al.~\cite{Cheng2022} construct~a knowledge graph~for~the \textit{Library} and \textit{Runtime} layers for general Python programs, but fail to support lower layers in the DL stack.}



Second, \textit{dependency recommendation techniques are needed}. Our introducing stage analysis reveals that environment~setup~is the~most bug-prone stage which introduces \cmmnt{\todo{87.4\%}~}\todob{90.8\%}~of~the~DBs. Therefore, developers often face difficulties in setting~up~a~feasible DL stack.~Further, our root cause analysis shows~that~\cmmnt{\todo{76.1\%}~}\todob{70.0\%} of the DBs~are~caused by \textit{Incompatible Software Version},~although dependency documentation provides prerequisite information about setting~up~dependencies and their version~constraints. Therefore, developers might not always refer to the documentation. In that sense, dependency~recommendation~techniques become useful for developers to ease~the~setup of~a~feasible DL stack; i.e., given some dependencies installed, they recommend other dependencies to form a complete DL stack. For example, given the available hardware and OS, they suggest required dependencies in \textit{Driver}, \textit{Runtime} and \textit{Library} layers.



Third, \textit{DB detection, localization and fixing techniques~are needed.} Our study indicates that \cmmnt{\todo{87.4\%}~}\todob{90.8\%} of the DBs~are~introduced in environment setup, while only \cmmnt{\todo{25.8\%}~}\todob{37.7\%} of the~DBs~are exposed in environment setup. Thus, \todoq{it may indicate that} many DBs stay undetected until later lifecycle stages. To~detect or localize~DBs~as~early~as possible, one possible remedy~is to identify the dependencies currently adopted in the DL stack, and then check against our dependency knowledge graph to detect~potential dependency constraint violations. Here~the~challenge~is~to~automatically identify all heterogeneous dependencies as well as their versions across the entire DL stack. Along this direction,~Tan~et al.~\cite{tan2022} proposed a technique to identify homogeneous dependencies at the \textit{Application} and \textit{Library} layer. Moreover,~as~many DBs are caused by software bugs or API incompatibilities, fine-grained call graph analysis is needed to accurately detect and localize DBs, i.e., to decide whether such bugs or incompatible APIs are~in~the execution path and thus can be triggered. Besides, \todohs{our study indicates that questioners are often unaware of the introducing dependencies of the DBs, which calls for automated DB localization techniques.} Once a DB is localized, automated fixing techniques can use the fix patterns derived from our study to fix it. However, the challenge is to decide which fix pattern or combination of fix patterns~is~applicable and how~a fix pattern is instantiated. A way is to use~search-based approach by applying~fix~patterns to generate potential fixes and using the dependency knowledge graph to decide the fix fitness.

Fourth, \textit{dependency upgrading and migration techniques are needed.} Our introducing stage analysis uncovers that~some~DBs are introduced in deployment and maintenance. More specifically, DL~stack in deployment environment can~be~different from the one in development environment. Hence, dependency migration techniques are needed to check whether dependencies in development environment can be replaced with~the~ones in deployment environment. Besides, dependency versions can be upgraded~for the benefit of fixed~bugs~and~improved features. However, it may also introduce incompatibilities. Therefore, dependency upgrading techniques are needed to analyze API changes and assess the risk in terms~of~potential DBs and the effort in terms of potential code adaptation.


\subsection{Application for Usefulness Demonstration}

To demonstrate the usefulness of our implications,~we~design a prototype to automatically detect and fix DBs. 

\textbf{Prototype Design.} Our prototype has one knowledge base, i.e., \textit{dependency constraint knowledge}, and two components,~i.e.,~\textit{DB~detection}~and \textit{DB fixing}. To collect dependency constraint knowledge, we target~at~the documentations of TensorFlow, Pytorch~and~Keras as~i)~they~expose and introduce the most DBs at the \textit{Library}~layer; and~ii) their documentations often list requirements for dependencies at lower layers, e.g., Python at the \textit{Runtime}~layer,~CUDA and cuDNN at the \textit{Driver} layer, Linux at the \textit{OS/Container} layer.~We~then manually extract dependency constraints from~their online documentations via reading installation guides~of~each~version, where they are either described in natural language or illustrated with a table. Each dependency constraint is denoted as a tuple $\langle dep_a, dep_b, v_a,$ $v_{b1} ,v_{b_2} \rangle$ where version $v_a$ of dependency $dep_a$ depends on~$dep_b$~under the condition that the version of $dep_b$ is within the range of $v_{b1}$ and $v_{b_2}$. $v_{b_2}$ can be null to represent an opening scope. Overall, we collect \todo{588} dependency constraints in \todo{3} days.

Our prototype takes a Docker image as an input,~and~detects~whether it contains a DB. If yes, it also tries to fix it. To detect DBs, we first need~to~identify~dependencies used in the DL stack. To this end, we support~two~package~managers, i.e., PyPI and Conda,~at~the~\textit{Library} layer. We obtain the virtual environment location of PyPI virtualenv by \textit{``find~/~| grep bin/activate''}, or the environment names of Conda~by~\textit{``conda env list''}. Then, we activate the corresponding~environments~by \textit{``source <path>/bin/activate''} or  \textit{``conda activate <envname>''}, where we retrieve the whole list of dependency versions under each environment. For the \textit{Runtime} layer, we identify Python runtime in PATH (\textit{``echo \$PATH''}), where there exists~an~executable named \textit{``python''} or  \textit{``python3''}. For the \textit{Driver} layer, we use \textit{``nvcc --version''} to identify installed version~of CUDA and \textit{``which nvcc''} to identify installed path of CUDA. Under \textit{``<cuda\_path>/lib64''}, we find the cuDNN version by checking if there exists a dynamic linking library of cuDNN (i.e., \textit{cudnn.so.<version>}). For the \textit{OS/Container} layer,~we~use \textit{``uname -a''}, \textit{``cat /etc/centos-release''}, \textit{``lsb\_release -a''}, etc. to identify hosting OS. We do not identify hardware versions as the Docker image does not contain hardware information.


Then, we search the~identified~dependency versions~of~each~environment~for~a~combination of dependencies $\langle dep_a, dep_b, v_a, v_{b}\rangle $ that violates a dependency constraint $\langle dep_a, dep_b, v_a, v_{b1}, v_{b2}\rangle $, which is regarded as a DB. \todohs{We first anchor $dep_a$ at version $v_a$ and change $dep_b$'s version indicated in version constraint from $v_{b1}$ to $v_{b2}$. Meanwhile, as $dep_b$'s version changes, we also change $dep_c$'s version to satisfy version constraint in $\langle dep_b, dep_c, v_b, v_{c1}, v_{c2}\rangle $ if needed. The process is conducted recursively. If there is no satisfied version combinations, we anchor $dep_a$'s version into a newer version~of~$v_a$~using our knowledge base and repeat the above process. Once a satisfied version is found, we change the dependency version by running uninstall and install script to fix the DB.}




\todoq{\textbf{Comparison with Related Tools.} We find and review three closely related tools. DockerizeMe~\cite{horton2019dockerizeme} is a tool to infer the dependencies needed to execute a Python code snippet without import error. The inference is based on a knowledge base which contains packages, their versions and resources, and the relationships between them. The knowledge base is built by applying static and dynamic analysis to top ten thousand Python packages on PyPI and applying association rule mining to public GitHub Python projects. PyEGo~\cite{Ye2022} extends the knowledge base of DockerizeMe by further including Python interpreters and system libraries, and achieves a better accuracy on inferring compatible dependencies. DockerizeMe and PyEGo mainly support packages installed by commands of \textit{pip} and \textit{apt}.
Different from DockerizeMe and PyEGo, our prototype extracts dependencies and version constraints knowledge from official documentation, and supports package installation commands beyond \textit{pip} and \textit{apt}. While further work is needed to automate the knowledge extraction, our approach can offer more generalizability across different types of dependencies at different DL stack layers.}

\todoq{Different from the knowledge-based inference in DockerizeMe and PyEGo, PyDFix~\cite{mukherjee2021fixing} takes a trial and error approach, i.e., it first identifies dependency errors and possible dependencies causing the~errors from build log, and then iteratively re-runs the build~with intermediate patches until the error disappears. Differently,~our~prototype does not rely on error logs since not all DBs indicate explicit error logs or reveal dependency names in their error logs.}

\textbf{Effectiveness Evaluation.} To evaluate our prototype,~we~reproduce DBs from our study and export them as Docker~images. 
We~randomly~sample \todo{80} DBs from our study, and successfully reproduce~\todo{18}~DBs. The reasons of unsuccessful reproduction~are~two-fold. First, the exposing or introducing dependency of DBs locates in~\textit{Hardware} or \textit{OS/Container} which does not match with~our~machines. Second, only part of the DL stack is revealed in the posts~or issues, and hence we fail to derive the full DL stack to reproduce~DBs. 

\todohs{Our prototype successfully detects and fixes 8 of the 18 DBs.~Three DBs caused~by \textit{Mismatched Software}, two DBs caused by \textit{Buggy~Software~Version} and one DB caused by \textit{Unsuccessful Installation} are not detected as our prototype is focused on violated version constraints. Of the twelve~DBs~caused~by~\textit{Incompatible Software Version}, five DBs are detected and fixed~using version constraint between TensorFlow and CUDA, two are detected and fixed using version constraint between CUDA and cuDNN, and one~is detected and fixed using version constraint between TensorFlow and cuDNN. The other four DBs are not detected because the root cause dependencies are not in our scope of dependency constraint knowledge acquisition. These results demonstrate the potential of our prototype.}


\todoq{Moreover, we try to apply PyEGo and PyDFix to fix the~18~DBs. Notice that DockerizeMe is not selected because PyEGo has achieved better performance than it.}
\todoq{We successfully run PyEGo against the 18 DBs. It successfully detects and fixes only one DB. It successfully detects 11 DBs, but generates wrong version recommendation on all of them. 
Besides, it fails to detect the rest 6 DBs. }
\todoq{Unfortunately,~we fail to launch PyDFix due to the limited setup documentation. However, PyDFix relies on analyzing error logs to fix DBs. Consequently, we can conclude that PyDFix are unable to fix at least 13 of the DBs since these DBs produce normal outputs instead of error~logs.~These results indicate the potential of our prototype.}

\todohs{\textbf{Human Study.} We observe from our effectiveness evaluation that our prototype takes \todo{2} seconds for the DBs that are not~successfully fixed, and these DBs are even not detected by our prototype. In other words, it takes negligible time for our prototype to determine whether a given DB is out of the scope of our prototype. Therefore, we are interested to investigate how much effort can be saved for developers for the DBs that are in the scope of our prototype.}

{To this end, we conduct a human study with \todo{8} participants to manually fix the \todo{8} DBs that can be automatically fixed by our prototype. The participants are recruited voluntarily at our college~who are familiar with Linux shell and packages, and has sufficient background in deep learning. \todoq{Four participants have worked on at least one or two research projects that employ DL techniques, and the other four participants have hands-on experience with open source DL projects.} The tasks are \todo{8} reproduced DBs in a Docker environment where the error trace of each DB could be invoked via a command (i.e., \textit{python script.py}). The participants are told that the error is caused by a DB and they are required to locate and fix the DBs with their expertise and any online resources. The order of the tasks are randomized for each participant to avoid bias.}

\todohs{We use two indicators to compare participants' manual fixes and our automated fixes. The first indicator is the \todo{quality} of the fix in each task. We use 2 to indicate a successful~and perfect fix, 1 to indicate a successful but imperfect fix, and~0~to~indicate an unsuccessful fix. The success of the fix is judged by the dismissing of the DB's errors when re-launching scripts. The perfection~and~imperfection of the fix is judged by two of the authors on whether the fix steps would have any side effect. After the discussion and mutual agreement from two of the authors, a final quality is resolved. The second indicator is the consumed time on finishing each task.}

\begin{figure}[!t]
   \centering
   \includegraphics[width=0.34\textwidth]{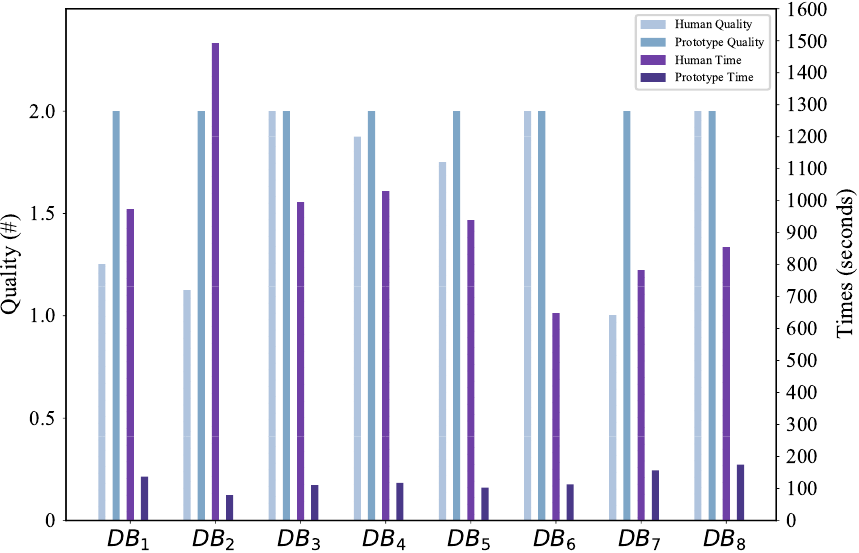}
   \vspace{-6pt}
   \caption{\todoq{Results of the Quality and Time of Fixing 8 DBs}}
   \label{fig:humanstudy}
\end{figure}

\todohs{Fig.~\ref{fig:humanstudy} shows the result of fix quality and time. In terms of quality, all 8 participants obtain full score in \todo{3} DBs. The rest \todo{5} DBs are~not fixed successfully and perfectly by all. \todo{19} participant-DB pairs are not fully scored. Specifically, we assign 1 to \tododel{5}\todoq{17} participant-DB pairs. Of these \tododel{5}\todoq{17} participant-DB pairs, \todo{2} participant-DB pairs fix a DB~by using soft links to redirect the incorrect dependency into a correct dependency and \todo{3} participant-DB pairs fix a DB by replacing dynamic linked libraries (i.e., change a correct dependency's file name into the original one using \textit{mv}). They are imperfect because such tricks are unstable and confuse other users. \todoq{The rest \todo{12} participant-DB pairs freshly reinstall TensorFlow using an up-to-dated version. We assign them to 1 because setting up a new environment carries the risk of disrupting the initial environment, making it impractical when there are multiple users and applications.} \tododel{It violates our aim for fixing the DBs in the experiment because instead of fixing the DBs, the participants.} We also assign~0~to~\tododel{14}\todo{2} participant-DB pairs. They fail to fix as it still has the reported error. In terms of time, none of the manual fix from \todo{64} participant-DB pairs surpasses our prototype. The manual fix takes averagely \todo{8.8} times longer than our prototype. Generally, our prototype achieves a higher quality of 2 against the human group with a score of 1.4, and costs averagely 109.2 seconds against the human group with averagely 963.0 seconds. Therefore, our prototype can be useful~for developers to provide high quality fix and greatly saving fixing~time.}

\section{Related Work}


\textbf{Dependency Bugs.} Dependency bugs have been explored~for~different ecosystems, e.g., Debian~and Red Hat~\cite{artho2012software},~JavaScript \cite{patra2018conflictjs},~Java \cite{wang2018dependency, wang2019could, wang2021will, huang2020interactive, macho2018automatically}, Python \cite{wang2020watchman, mukherjee2021fixing}, C/C++~\cite{jia2021depowl}~and Go \cite{wang2021hero}. 
\todo{To~the best of our knowledge,~our work is the first to systematically investigate dependency bugs in DL ecosystem.} 

\textbf{Deep Learning Bugs.} Empirical studies have been conducted~to characterize bugs in DL systems. Some are focused~on~a~general~scope of bugs \cite{EmpiricalStudyTensorFlow2018, islam2019comprehensive, humbatova2019taxonomy, islam2020repairing, nikanjam2022faults}, and others are focused~on~a~specific type of bugs \cite{zhang2020empirical, chen2021empirical, zhang2021autotrainer, MLAPI2021, cao2021characterizing}. 
\todo{These studies uncover partial characteristics~of~dependency~bugs in DL stack.~There~lacks~a comprehensive study~to~characterize dependency bugs in DL stack,~and our work fills~this~gap.} Several advances have also~been made~to~detect DL bugs, e.g., numerical bugs \cite{zhang2020detecting, wardat2021deeplocalize,yan2021exposing} and shape bugs~\cite{lagouvardos2020static, Verma2020, liu2021detecting, wu2021tensfa}. \todo{However, little attention has been received to detecting dependency bugs in DL stack, and our work sheds~light~on it.}

\textbf{Empirical Studies about DL.} Many studies have empirically investigated various aspects~in~developing, deploying~and~maintaining DL systems \cite{giray2021software, martinez2022software, amershi2019software, alshangiti2019developing, zhang2019empirical, chen2020comprehensive, Cummaudo2020, pham2020problems, ma2019moving, guo2019empirical, tang2021empirical, nikanjam2021design, Sculley2015} and DL frameworks \cite{han2020programmers, zhang2021unveiling, liu2021exploratory, liu2020using, han2020empirical, tan2022}.~\todo{These~studies motivate~the importance of dependency management. ~For~example, incompatible dependency installation or environment setup~is~recognized~as~a~common challenge \cite{chen2020comprehensive, zhang2019empirical, alshangiti2019developing}. 
However, they lack~an~in-depth analysis of the characteristics.~Our~work~is inspired by them to systematically characterize dependency bugs across the DL~stack.}

\section{Conclusions}

We have conducted the first comprehensive study~to~characterize DBs~across the entire DL stack. We~provide useful findings to raise the awareness of DBs in DL stack in the DL community,~and~provide actionable implications for developers and researches.


\section{Data Availability}

The data of our study is available at \todo{https://dl-dep.github.io}.

\begin{acks}
This work was supported by the National Key R\&D Program~of~China (2021ZD0112903) and the China Postdoctoral Science Foundation (2022M720768).
\end{acks}


\bibliographystyle{ACM-Reference-Format}
\bibliography{src/reference,src/so}

\end{document}